\documentclass{zhilingResearch}

\usepackage{enumitem}
\usepackage{url}
\usepackage{indentfirst}
\usepackage{float}
\usepackage{threeparttable}
\usepackage{array}
\usepackage[section]{placeins}
\renewcommand{\thefootnote}{\arabic{footnote}}

\title{Ozone: A Unified Platform for Transportation Research}

\author
{Ou Zheng~$^{a, b}$,  Ruyi Feng~$^{a}$\begingroup\renewcommand{\thefootnote}{1}\footnote{Correspondence E-mail: fengruyi@seu.edu.cn}\endgroup,Yufeng Yang~$^{a}$, Shengxuan Ding~$^{a}$, Lishengsa Yue~$^{c}$, Ye Li~$^{d}$, Baichuan Mo $^{e}$, Yunhan Zheng~$^{b}$, Minwei Kong~$^{f}$, Dingyi Zhuang~$^{f}$, Ao Qu~$^{f}$, Zhibin Li~$^{g}$, Meng Li~$^{h}$, Dongjie Wang~$^{i}$, Wangyang Ying~$^{a}$  \\
\vspace{1em}
\normalfont{\small $^{a}$Zhiling Research}\\
\normalfont{\small $^{b}$Peking University}\\
\normalfont{\small $^{c}$Tongji University}\\
\normalfont{\small $^{d}$Central South University}\\
\normalfont{\small $^{e}$Tsinghua University}\\
\normalfont{\small $^{f}$Massachusetts Institute of Technology}\\
\normalfont{\small $^{g}$Southeastern University}\\
\normalfont{\small $^{h}$Korea Advanced Institute of Science and Technology }\\
\normalfont{\small $^{i}$University of Kansas }\\
}

\begin{document}

\maketitle
\thispagestyle{firstpagestyle}

\begin{abstract}
Transportation research increasingly depends on heterogeneous data from roadside cameras, UAV imagery, LiDAR, and in-vehicle sensors. Differences in data representations, model interfaces, and evaluation protocols complicate cross-dataset benchmarking and reproducibility. We present Ozone, a unified transportation research platform organized around Hardware, Data, Model, and Evaluation layers, with a Prototype layer supporting deployment-oriented applications. Its data layer links vehicle trajectories, traffic conflicts, reconstructed crashes, and digital-twin maps through a shared trajectory schema. The source inventory spans 19 trajectory datasets recorded at 59 sites in eight countries, with completed converters for 12 sources and adaptation ongoing for seven. The converters align coordinate systems, object classes, oriented bounding boxes, and kinematic fields. A shared implementation of surrogate safety measures supports conflict extraction, while map and motion constraints link crash reports to reconstructed trajectories. CARLA digital-twin maps and SUMO traffic logic connect these records to simulation, model development, and evaluation. Validation on 500 vehicles from four datasets reports mean position, heading, and speed errors below 0.15\,m, 2.5$^\circ$, and 0.3\,m/s. A workflow study reports an 85\% reduction in experiment setup time. Case studies demonstrate human-factor experiments and automated-driving safety testing, while the model interfaces support traffic modeling and world-model research. The source code and datasets are publicly available.\begingroup\renewcommand{\thefootnote}{2}\footnote{\url{https://ozone.zhilingtech.com/}}\endgroup

\end{abstract}

\section{Introduction}

Intelligent transportation systems increasingly rely on multi-sourced sensing, including roadside cameras, in-vehicle cameras, drone imagery, radar, and LiDAR \citep{zheng2023citysim, krajewski2018highd}. These sensors generate heterogeneous data covering traffic-control logs, road and infrastructure attributes, vehicle trajectories, and human-centered signals such as driver state and physiological measurements \citep{liu2025eeg}. At the same time, automated driving and emerging traffic foundation models scale up their dependence on multi-modal evidence, because they must combine on-board perception with bird's-eye and roadside views to interpret complex, interactive urban scenes \citep{feng2023dense, russell2025gaia}. As transportation data become more multi-modal and more widely distributed across platforms, the central challenge is now consistency of integration rather than volume of collection. Joint analysis requires the alignment of spatial coordinates, temporal synchronization, and semantic labels across vehicle-side and infrastructure-side observations. Without that alignment, data from different sources cannot be reliably fused for modeling, evaluation, or deployment \citep{tao2019digital}.

Existing datasets follow different standards. Widely used trajectory datasets, including NGSIM \citep{ngsim2016}, highD \citep{krajewski2018highd}, CitySim \citep{zheng2023citysim}, and UTE \citep{feng2025drone}, each adopt their own coordinate systems, sampling rates, object representations, and metadata conventions. Many omit the motion fields that safety analysis requires, including heading and the oriented bounding box \citep{arun2021systematic}. A study that draws on two of them must reconcile these choices by hand, and a study that transfers a model from one city to another must first establish that the inputs mean the same thing in both. Safety evidence is fragmented further. Conflicts are derived quantities whose definitions and thresholds vary across studies, so their event counts are difficult to compare directly. Crash reports, which remain the primary record of severe outcomes, describe discrete facts rather than continuous motion, and therefore cannot be placed beside a trajectory without first recovering the motion the report never contained \citep{wang2019crash, kumar2018novel}.

Simulation adds a third layer of inconsistency. Platforms such as SUMO and CARLA expect their own map and scenario descriptions \citep{lopez2018sumo, dosovitskiy2017carla}, and public trajectory datasets seldom ship a matched map, so every group that needs one rebuilds it. Differences between independently reconstructed roads can affect results even when studies nominally use the same site. Assembling an experimental stack that spans sensing, data, and simulation therefore takes months of engineering before a new question can be asked, which slows methodological progress and weakens reproducibility \citep{fuller2020digital}.

We propose \textbf{Ozone}, a unified transportation research platform that connects sensing, standardized data, models, evaluation, and deployment-oriented prototypes. Four core research layers organize the scientific workflow, and a fifth Prototype layer extends it toward application. The data layer provides a shared representation of vehicle trajectories, traffic conflicts, reconstructed crashes, and digital-twin maps. The model and evaluation layers build on these records through common interfaces, benchmark configurations, and metric implementations. This broader platform scope supports both dataset reuse and the experiments that depend on it.

The first source inventory covers 19 trajectory datasets at 59 sites in eight countries. Converters are complete for 12 sources, while seven remain under adaptation. The shared schema preserves source and site information while aligning spatial coordinates, object representations, and motion fields. Conflicts are extracted from these standardized tracks with a common implementation of surrogate safety measures. Crash reports are linked to physically plausible reconstructed trajectories, and matched digital-twin maps support replay, additional sensor viewpoints, and closed-loop simulation. These resources also form the basis of the Next-generation Benchmark by Digital Twin (NBDT), the scenario-oriented benchmark component of Ozone.

The main contributions are as follows.
\begin{itemize}
    \item We organize transportation research into four core layers and a Prototype extension, with schemas and interfaces connecting data collection to deployment-oriented validation.
    \item We integrate trajectories, traffic conflicts, reconstructed crashes, and digital-twin maps under a shared data schema, with source-specific converters and explicit source and site metadata.
    \item We connect data conversion and raw-sensor processing to model interfaces and common evaluation protocols, supporting traffic modeling, safety analysis, human-factor experiments, and world-model research.
    \item We assess conversion accuracy and workflow efficiency, and demonstrate the platform through emergency-takeover experiments and automated-driving tests involving road-engineering hazards.
\end{itemize}

The remainder of the paper first defines the research layers, then presents the dataset and its processing procedures. It next describes the model, evaluation, and prototype interfaces, followed by case studies, system evaluation, and future work.


\section{Transportation Research Layers}\label{sec:layers}

\begin{figure}[tbp]
\centering
\includegraphics[width=\linewidth]{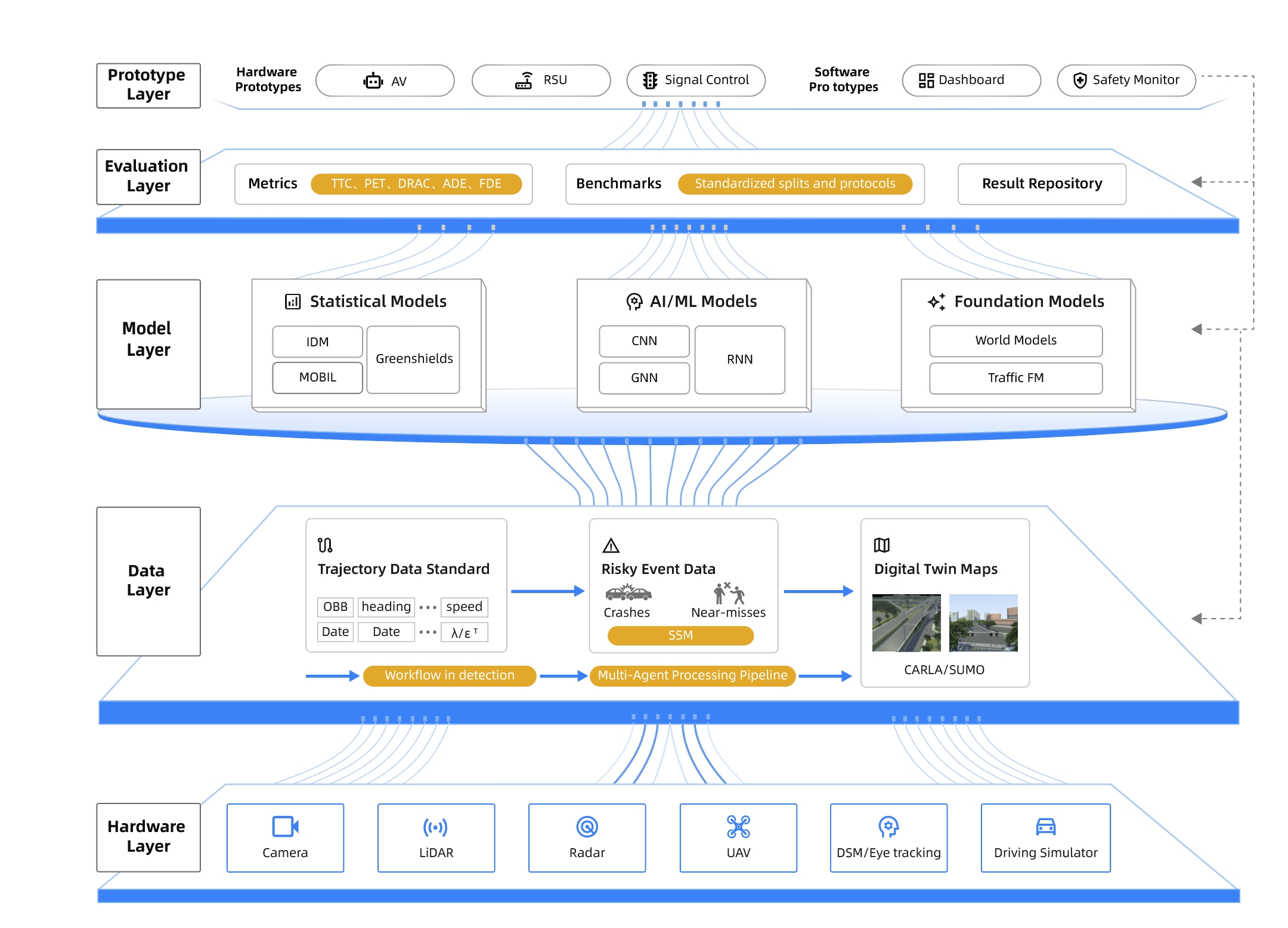}
\caption{Ozone transportation research framework, with four core research layers and a Prototype extension.}
\label{fig:framework}
\end{figure}

Ozone organizes the transportation research workflow into four core layers, covering Hardware, Data, Model, and Evaluation. A fifth Prototype layer extends this workflow toward deployment. Figure~\ref{fig:framework} shows their relationships. The hardware layer supplies observations and experimental platforms. The data layer converts these observations into shared records and map assets. The model layer uses those records for prediction, simulation, and safety analysis. The evaluation layer defines how the outputs are assessed, and the prototype layer connects validated methods to applications.

\subsection{Hardware Layer}

The hardware layer encompasses all physical devices and simulation platforms used to observe and recreate transportation phenomena. Traffic data collection relies on several sensor types. Inductive loop detectors provide aggregate flow counts, fixed roadside cameras capture intersection-level video, UAVs deliver bird's-eye-view imagery over extended areas, and vehicle-mounted LiDAR and radar supply high-resolution three-dimensional point clouds \citep{zheng2023citysim}. In-vehicle sensors further expand this spectrum to include GPS/INS for ego-vehicle positioning, dashboard cameras for forward-scene capture, and physiological instruments such as electroencephalography (EEG) and eye-tracking devices for driver-state monitoring \citep{liu2025eeg, al-gburi2024driver}.

Beyond physical sensing hardware, the hardware layer also includes simulation platforms that replicate real-world conditions in a controlled environment. Driving simulators enable human-in-the-loop experiments with configurable road geometries, traffic densities, and weather conditions, providing repeatable scenarios for human-factor studies. High-fidelity co-simulation frameworks that couple microscopic traffic simulators (e.g., SUMO) with rendering engines (e.g., CARLA) support autonomous-vehicle testing with realistic perception pipelines \citep{lopez2018sumo, dosovitskiy2017carla}.

\subsubsection{Problem Statement}

The core challenge at the hardware layer is \emph{sensor heterogeneity}. Each device type produces data in its own coordinate frame, resolution, and temporal cadence. A roadside camera operates at 30\,fps in pixel coordinates, whereas a LiDAR scanner outputs 10\,Hz point clouds in a local Cartesian frame. Fusing these streams into a coherent spatiotemporal representation requires explicit calibration metadata. This includes intrinsic and extrinsic parameters, synchronization timestamps, and georeferencing anchors, which most current datasets do not standardize. Ozone addresses this gap by defining a hardware schema that records sensor type, mounting geometry, coordinate reference, and sampling configuration alongside every data record.

\subsection{Data Layer}

The data layer concerns how traffic facts are measured, stored, and curated. It covers three principal data categories, comprising (i) trajectory data that record individual road users' motion as time-ordered position sequences, (ii) event data that capture safety-critical incidents such as crashes and near-misses, and (iii) contextual data that provide environmental attributes including weather, lighting, and signal-timing logs.

Trajectory datasets form the backbone of transportation research. From trajectory maps, researchers can extract macroscopic traffic-flow parameters such as average speed, density, and volume, which are traditionally obtained from inductive loop detectors \citep{coifman2015empirical}. Trajectory data also offer an intuitive view of traffic phenomena such as kinematic waves and traffic breakdowns, as well as microscopic parameters including individual vehicle speeds, time headways, and space headways. Numerous studies have utilized vehicle trajectory data for traffic-flow model calibration \citep{talebpour2015influence}, traffic feature exploration \citep{wan2020spatiotemporal}, and car-following and lane-changing behavior analysis \citep{treiber2000congested, kesting2007mobil}.

\subsubsection{Data Collection}

Trajectory data can be collected through multiple sensing modalities. Roadside cameras provide a fixed field of view over intersections, whereas UAV-based imaging offers flexible, wide-area coverage of highway segments, weaving zones, and merge areas \citep{zheng2023citysim, berghaus2024vehicle}. Connected-vehicle technologies and GPS probe data add another source of trajectory information, though typically at lower spatial precision. Each collection method introduces characteristic noise patterns, occlusion profiles, and coverage limitations that downstream processing must accommodate.

\subsubsection{Data Standardization}

A trajectory standard that unifies coordinate definitions, timestamp conventions, object representations, and metadata conventions across datasets can substantially reduce friction in data reuse and cross-dataset evaluation. The minimal fields required include a unique vehicle identifier, frame-level timestamps, position in a calibrated local coordinate system, oriented bounding box (OBB) corners, heading angle, instantaneous speed, and two-dimensional acceleration components. When these fields are consistently available, derived quantities such as surrogate safety measures, fundamental-diagram parameters, and car-following model inputs can be computed consistently across studies.

\subsubsection{Problem Statement}

Currently, widely used datasets such as NGSIM \citep{ngsim2016}, highD \citep{krajewski2018highd}, CitySim \citep{zheng2023citysim}, and UTE \citep{feng2025drone} each follow their own conventions for coordinate systems, bounding-box representations, and metadata fields. Many omit essential kinematic variables (e.g., heading, OBB corners) that are critical for accurate safety-metric computation \citep{arun2021systematic}. This inconsistency forces researchers to write dataset-specific parsers and conversion scripts, duplicating effort and introducing potential errors. The data layer of Ozone defines a canonical schema and provides automated conversion pipelines that map each source dataset to this schema, filling in missing fields through geometric and kinematic inference.

\subsection{Model Layer}

The model layer spans the computational methods applied to transportation data, covering the full spectrum from classical statistical and physics-informed models to machine-learning, deep-learning, and emerging foundation models. It serves as the core analytical layer that transforms standardized transportation data into prediction, simulation, safety assessment, scene generation, and decision-support capabilities, thereby linking data resources with benchmarked research outputs and downstream applications.

\subsubsection{Statistical Models}

Traditional approaches to traffic analysis employ parametric models grounded in traffic-flow theory. Fundamental-diagram models, including the Greenshields linear model, Greenberg's logarithmic model, and the Underwood exponential model, relate macroscopic variables (flow, density, speed) and serve as calibration targets for simulation environments \citep{greenshields1935study}. At the microscopic level, car-following models such as the Intelligent Driver Model (IDM) \citep{treiber2000congested} and lane-changing models such as MOBIL \citep{kesting2007mobil} describe individual vehicle interactions and form the basis for traffic microsimulation.

\subsubsection{Machine Learning and Deep Learning Models}

Machine-learning methods extend the modeling palette beyond physics-based formulations. Gradient-boosted trees and support-vector machines have been applied to crash-risk prediction and traffic-state estimation, while deep-learning architectures have advanced trajectory prediction, anomaly detection, and traffic signal optimization. These include convolutional networks for spatial features, recurrent networks for temporal sequences, and graph neural networks for relationships between road users or network elements \citep{wu2024accidentgpt}.

\subsubsection{Foundation Models}

An emerging class of large-scale foundation models is being developed specifically for transportation applications. World models for autonomous driving learn environment dynamics from video data and can generate plausible future traffic scenes, supporting planning and prediction without extensive on-road data collection \citep{russell2025gaia, zheng2025world}. Traffic foundation models aim to provide generalizable representations that transfer across tasks such as flow prediction, incident detection, and signal control, analogous to how language models generalize across natural-language tasks.

\subsubsection{Problem Statement}

The model layer faces two related challenges. Inconsistent input and output interfaces make it difficult to replace a model within an experimental pipeline. Different evaluation protocols also limit fair comparison across studies. Ozone addresses these by defining model schemas that specify input formats, output formats, and evaluation hooks, enabling plug-and-play model integration.

\subsection{Evaluation Layer}

The evaluation layer defines how research outcomes are measured, benchmarked, and compared across tasks and application settings. It provides the criteria that connect models and systems to scientific claims, translating experimental outputs into standardized evidence on accuracy, safety, robustness, transferability, and practical effectiveness, thereby enabling fair comparison and reproducible validation across transportation research domains.

\subsubsection{Standard Metrics}

Surrogate safety measures (SSMs) provide quantitative indicators of conflict severity without requiring actual crash data. Widely used SSMs include Time to Collision (TTC), Post-Encroachment Time (PET), Time Exposed TTC (TET), Time Integrated TTC (TIT), Deceleration Rate to Avoid a Crash (DRAC), and the Collision Avoidance Index (CAI) \citep{arun2021systematic, singh2024conflict}. Statistical criteria such as root-mean-square error (RMSE), mean absolute error (MAE), and correlation coefficients evaluate predictive model accuracy.

\subsubsection{Benchmarks}

Benchmark protocols vary by research domain. Computer-vision benchmarks evaluate object-detection and tracking performance using metrics such as intersection-over-union (IoU) and multi-object tracking accuracy (MOTA). Driving-simulation benchmarks assess human-factor responses under controlled experimental conditions. Prediction benchmarks compare trajectory-forecasting models on standardized test splits with metrics such as average displacement error (ADE) and final displacement error (FDE) \citep{ramesh2025habit}. Autonomous-vehicle safety benchmarks measure crash rates, near-miss frequencies, and testing efficiency in simulation and field experiments \citep{feng2023dense, zhou2024diffroad}.

\subsubsection{Problem Statement}

Without standardized benchmark protocols, researchers often report results using different metrics, dataset splits, or evaluation windows, making direct comparison unreliable. Ozone addresses this by providing pre-defined evaluation suites for each research domain, including fixed dataset splits, metric computation scripts, and baseline result repositories.

\subsection{Prototype Layer}

The Prototype layer connects research outputs to hardware and software applications. Hardware examples include automated vehicles, roadside units, and adaptive signal controllers. Software examples include traffic-management dashboards and safety-monitoring services. Shared interfaces allow these applications to consume the same trajectory and event records used during model development. Deployment-oriented evaluation also considers operating constraints and monitoring requirements, as described in Section~\ref{sec:prototype}.


\section{The Ozone Dataset}\label{sec:dataset}

The data layer connects microscopic vehicle motion with traffic-flow analysis, safety assessment, and simulation. Ozone organizes four related resources under one schema, comprising vehicle trajectories, identified traffic conflicts, reconstructed crashes, and digital-twin maps. The trajectory is the central record. A conflict links an interaction window to the tracks of the involved road users. A reconstructed crash links report attributes to an inferred motion sequence. A digital-twin map provides the road geometry for replay and simulation. Shared scene metadata preserves the relationship between each record, its source, and its site.

\begin{figure}[tbp]
  \centering
  \includegraphics[width=\linewidth]{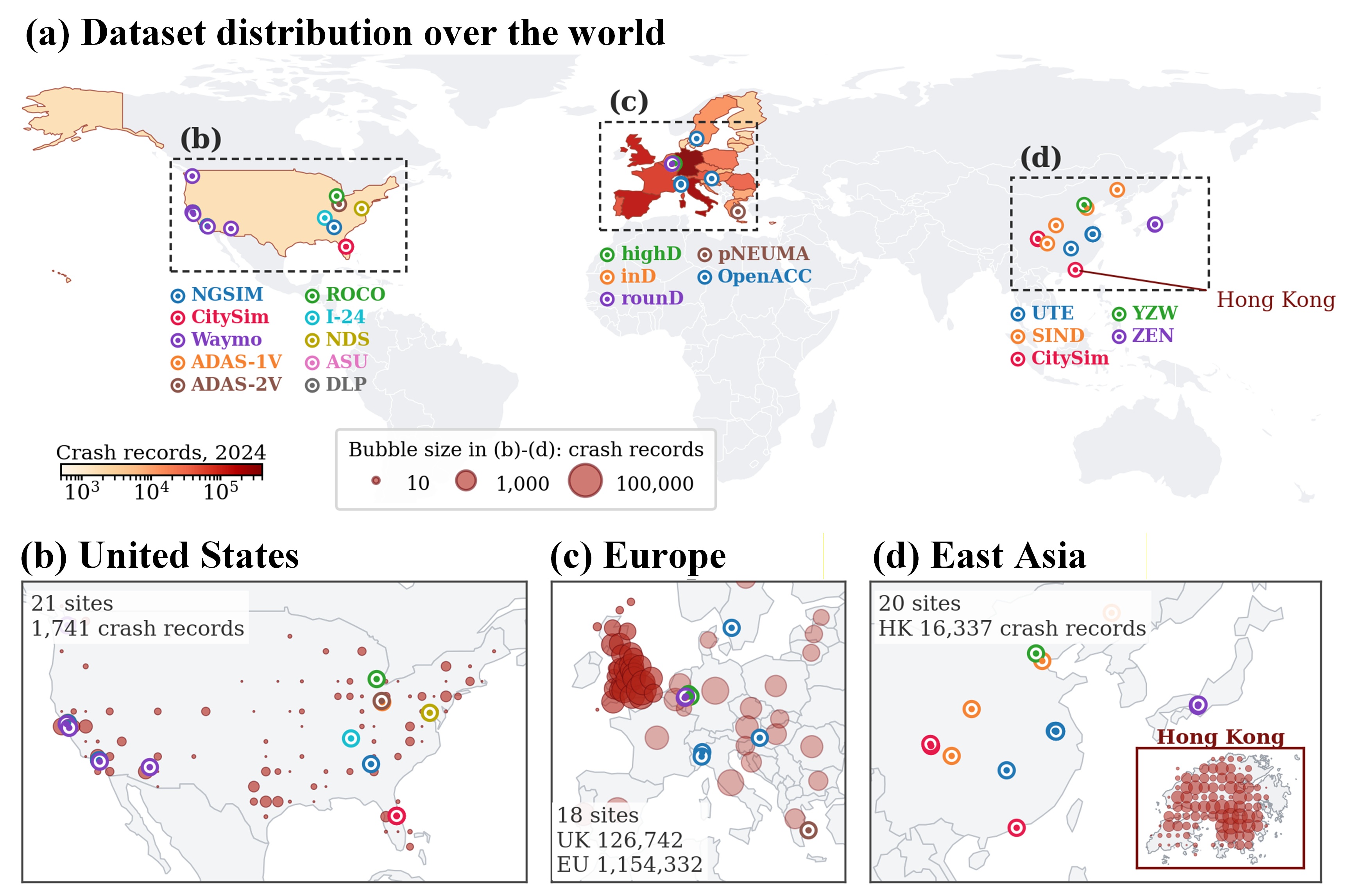}
  \caption{Geographic inventory of trajectory sources and crash-report coverage. Trajectory conversion is complete for 12 of the 19 sources.}
  \label{fig:worldmap}
\end{figure}

The source inventory and the converted release have different coverage. Figure~\ref{fig:worldmap} summarizes the geographic inventory, while Table~\ref{tab:datasets} lists the sources with completed converters. Counts of source records, converted trajectories, extracted conflicts, and reconstructed crashes therefore describe different resources and should be interpreted separately.

\subsection{Geographic Coverage and Data Sources}

The source inventory comprises 19 trajectory datasets recorded at 59 sites across the United States, Germany, China, Greece, Japan, Sweden, Italy, and Hungary. Twelve of these sources have completed converters and the remaining seven are being adapted to the shared representation. Panel (a) of Figure~\ref{fig:worldmap} places the sites on the world map, with the ring color identifying the source dataset, and panels (b) to (d) give the regional detail for the United States, Europe, and East Asia. The crash-source inventory extends the map further, shading the jurisdictions of the United States, the United Kingdom, the European Union, and Hong Kong by the crash records they contribute in 2024. In the regional panels the bubble radius scales with the logarithm of that count, aggregated for display using 150 km cells in the United States, major urban areas in the United Kingdom, whole countries in the European Union, and 3.5 km cells in Hong Kong. Reporting regimes differ, so these counts are not comparable across jurisdictions and the shading should be read as reach rather than as a rate. For converted sources, the shared schema removes source-format differences while retaining the original sensing and recording context. The geographic inventory does not imply that every mapped source has completed conversion or that every crash report has been reconstructed.

The sources behind this coverage draw on five sensing modalities. Drone video provides bird's-eye tracks over intersections, roundabouts, weaving zones, and merge areas \citep{zheng2023citysim, feng2025drone}. Fixed and roadside cameras provide long recordings over freeways, urban expressways, and arterials \citep{ngsim2016, zen2018ztd}. Roadside sensor installations provide continuous coverage of a single junction \citep{zhang2022design}. In-vehicle logs provide ego-centered motion together with the surrounding vehicles, and automated-vehicle sensor suites provide urban scenes \citep{sun2020scalability}. Satellite positioning provides platoon tracks from car-following experiments \citep{makridis2021openacc}.

Table~\ref{tab:datasets} lists the twelve sources whose converters are complete. Between them they cover freeways, urban expressways, signalized and unsignalized intersections, roundabouts, arterials, and merge and weaving segments, which are the road types that different research questions require. Each source is accompanied by a converter that reads the raw files and writes standardized tracks, so several datasets can be combined in one study without new parsers.

\begin{table}[tbp]
  \caption{Trajectory Sources with Completed Converters in the First Ozone Release}\label{tab:datasets}
  \begin{center}
  \footnotesize
  \begin{threeparttable}
  \begin{tabular}{@{}l l l >{\raggedright\arraybackslash}p{3.4cm} c@{}}
    \toprule
    Source & Region & Sensing & Road type & Rate \\
    \midrule
    NGSIM & USA & Fixed camera & Freeway, arterial & 10 \\
    highD & Germany & Drone & Highway & 25 \\
    inD & Germany & Drone & Intersection & 25 \\
    rounD & Germany & Drone & Roundabout & 25 \\
    CitySim & USA, China & Drone & Intersection, roundabout, expressway & 30 \\
    UTE & China & Drone & Expressway, merge, weaving & 30 \\
    ROCO & USA & Roadside sensors & Roundabout & 2.5 \\
    ZEN & Japan & Fixed camera & Urban expressway & 10 \\
    Waymo & USA & AV sensors & Urban & 10 \\
    ADAS single-vehicle & USA & In-vehicle & Arterial corridor & 10 \\
    ADAS two-vehicle & USA & In-vehicle & Arterial corridor & 10 \\
    OpenACC & SE, IT, HU & GNSS platoon & Highway, proving ground & 10 \\
    \bottomrule
  \end{tabular}
  \begin{tablenotes}[flushleft]
    \footnotesize
    \item[] \textit{Note.} Rate is the trajectory sampling frequency in Hz. Source timestamps are retained during conversion. SE, IT, and HU denote Sweden, Italy, and Hungary.
  \end{tablenotes}
  \end{threeparttable}
  \end{center}
\end{table}

The sources differ in scale and setting. CitySim contributes 13 recording sites and approximately 977 minutes of drone video at 30 frames per second, covering signalized and unsignalized intersections in Florida and expressway segments in China. The three German drone datasets add depth by road type, with 60 highway recordings in highD, 33 intersection recordings in inD, and 24 roundabout recordings in rounD \citep{krajewski2018highd, bock2020ind, krajewski2020round}. NGSIM contributes four classic freeway and arterial sites. UTE adds expressway sites in Nanjing recorded under free-flow, congested, and transitional conditions. ROCO adds sustained roadside observation of a single Michigan roundabout, which yields interactions between circulating and entering vehicles that drone recordings capture only in short windows \citep{zhang2022design}. ZEN adds trajectories from the Hanshin Expressway in Japan, the only urban expressway in the release recorded by fixed cameras over an extended corridor \citep{zen2018ztd}. The two driver-assistance collections from Central Ohio add 216 single-vehicle and 68 two-vehicle route runs, and they record the following-distance and driving-style settings of each run, which public data seldom do. OpenACC adds 139 car-following runs from proving grounds and public roads in Sweden, Italy, and Hungary \citep{makridis2021openacc}.

The platform resources total approximately 2.3\,TB. Once a source-specific converter is available, processing a dataset takes approximately four to eight hours on a single machine, depending on source size and field completeness. This runtime does not include the effort required to develop a new converter.

\subsection{Trajectory Schema}

Combining sources with this geographic and sensing coverage requires a common data format. Ozone takes the road-user track as its basic unit and defines the fields used for modeling, safety analysis, and simulation. Every road user in a scene is one track with a stable identifier, and at every frame the track stores the position of the vehicle in a local metric frame, the four corners of an oriented bounding box, the heading, the speed, and the two components of acceleration. Global coordinates in WGS84 are stored when the source provides them, which allows the site to be located on a world map and registered against its digital twin. Each track also carries an object class and a lane identity. Scene-level metadata records the source dataset, the site, the recording date and time, the frame rate, the weather, and the links to a two-dimensional basemap and a three-dimensional CARLA map. Table~\ref{tab:schema} lists the main fields.

\begin{table}[tbp]
  \caption{Main Fields of the Ozone Trajectory Schema}\label{tab:schema}
  \begin{center}
  \begin{tabular}{@{}p{6.0cm} >{\raggedright\arraybackslash}p{10.2cm}@{}}
    \toprule
    Field & Description \\
    \midrule
    datasetName, siteName & Source dataset and recording site \\
    carId & Unique track identifier for each road user \\
    frameNum, time & Frame index and timestamp \\
    carCenterXm, carCenterYm & Position in a local metric frame (m) \\
    carCenterLon, carCenterLat & Global position in WGS84, when available \\
    boundingBox (4 corners) & Oriented bounding box in metric and image frames \\
    heading & Orientation angle of the vehicle \\
    speed & Instantaneous speed \\
    accX, accY & Two-dimensional acceleration \\
    objClass & Road-user class (car, truck, bus, and others) \\
    laneId & Lane identity \\
    map, digitalTwin & Links to the basemap and the CARLA map \\
    \bottomrule
  \end{tabular}
  \end{center}
\end{table}

Vehicle shape and orientation determine the occupied road space and therefore affect safety measures \citep{arun2021systematic}. Ozone represents this geometry through oriented bounding boxes (OBBs) and headings. Missing fields are completed when the source provides sufficient geometric or kinematic information, following the procedure in Section~\ref{sec:conversion}. An inferred box remains an estimate, particularly when dimensions must be assigned from the object class. Sources without sufficient geometry are excluded from analyses that require vehicle dimensions.

The schema provides a common representation without implying a common sampling rate. The sources in Table~\ref{tab:datasets} retain their native timestamps and rates. For example, 25\,Hz and 30\,Hz correspond to intervals of 0.04\,s and approximately 0.033\,s. Any resampling needed for a particular model belongs to the benchmark configuration and must use a consistent rule across sources. WGS84 positions and image coordinates are retained when available, rather than inferred from an unreferenced local frame.

\subsection{Traffic Conflict Data}

Crashes rarely appear in ordinary trajectory recordings, while near-miss interactions provide evidence of the maneuvers and adjustments that precede unsafe outcomes. Ozone adds a conflict subset that captures the near-miss events contained within normal motion. Conflict extraction is performed from standardized trajectories, retaining only segments where road users approach collisions \citep{arun2021systematic, wang2024assessing}. Based on surrogate safety measures, risk labels can be generated without manual annotation.

\begin{figure}[tbp]
	\centering
	\includegraphics[width=0.9\linewidth]{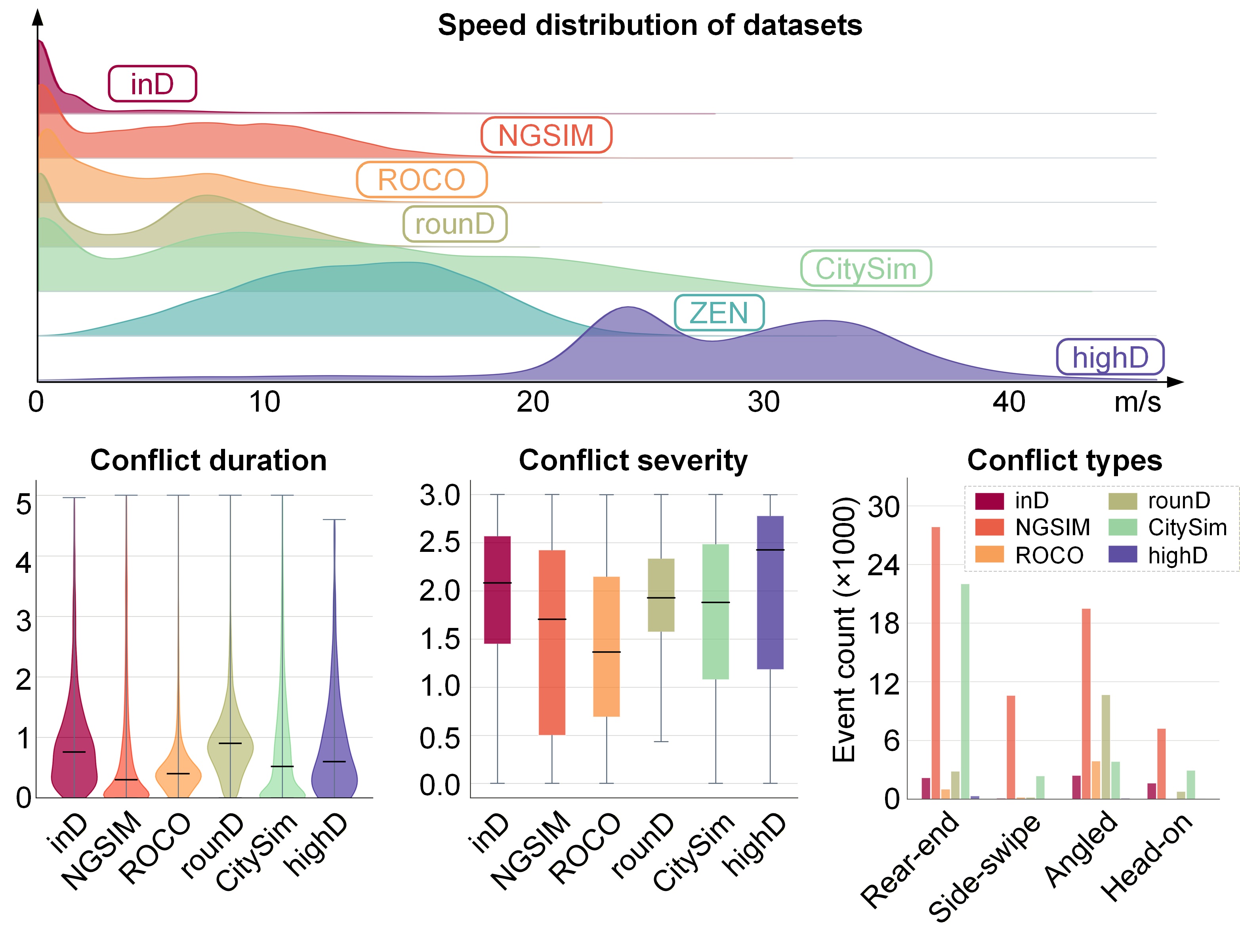}
	\caption{Speed distribution of seven Ozone sources, with the duration, severity, and type of the conflicts identified in them.}
	\label{fig:param}
\end{figure}

Supported surrogate safety measures are listed in Table~\ref{tab:ssm}. Each event record retains the involved tracks over the interaction window, the conflict type, a severity level, and temporal markers for onset, minimum TTC or PET, and resolution. Conflict types include rear-end, sideswipe, angle, and head-on interactions. A shared implementation makes the measure definitions consistent, while the selected thresholds and observation windows still determine which events enter an analysis. Figure~\ref{fig:param} summarizes the result for seven typical sources. The upper panel gives the speed distribution of each source, and the three lower panels give the duration, the severity, and the type of the conflicts identified in it. ZEN appears in the speed panel alone, because vehicle geometry is insufficient in the converted records used for this analysis. The seven sources still being adapted are also excluded from the conflict statistics. Conflict coverage is therefore narrower than the trajectory-source inventory.

\begin{table}[tbp]
  \caption{Surrogate Safety Measures Precomputed in Ozone}\label{tab:ssm}
  \begin{center}
  \small
  \begin{tabular}{@{}p{6.3cm} l >{\raggedright\arraybackslash}p{8.0cm}@{}}
    \toprule
    Measure & Short name & What it captures \\
    \midrule
    Time to collision & TTC & Time to collision under assumed motion \\
    Two-dimensional time to collision & 2D-TTC & TTC that accounts for both axes \\
    Post-encroachment time & PET & Gap between two users crossing a point \\
    Time exposed TTC & TET & How long a track stays below a TTC threshold \\
    Time integrated TTC & TIT & Integrated severity below a TTC threshold \\
    Deceleration rate to avoid a crash & DRAC & Required deceleration to avoid collision \\
    Collision avoidance index & CAI & Combined index of avoidance difficulty \\
    \bottomrule
  \end{tabular}
  \end{center}
\end{table}

\begin{figure}[tbp]
  \centering
  \includegraphics[width=\linewidth]{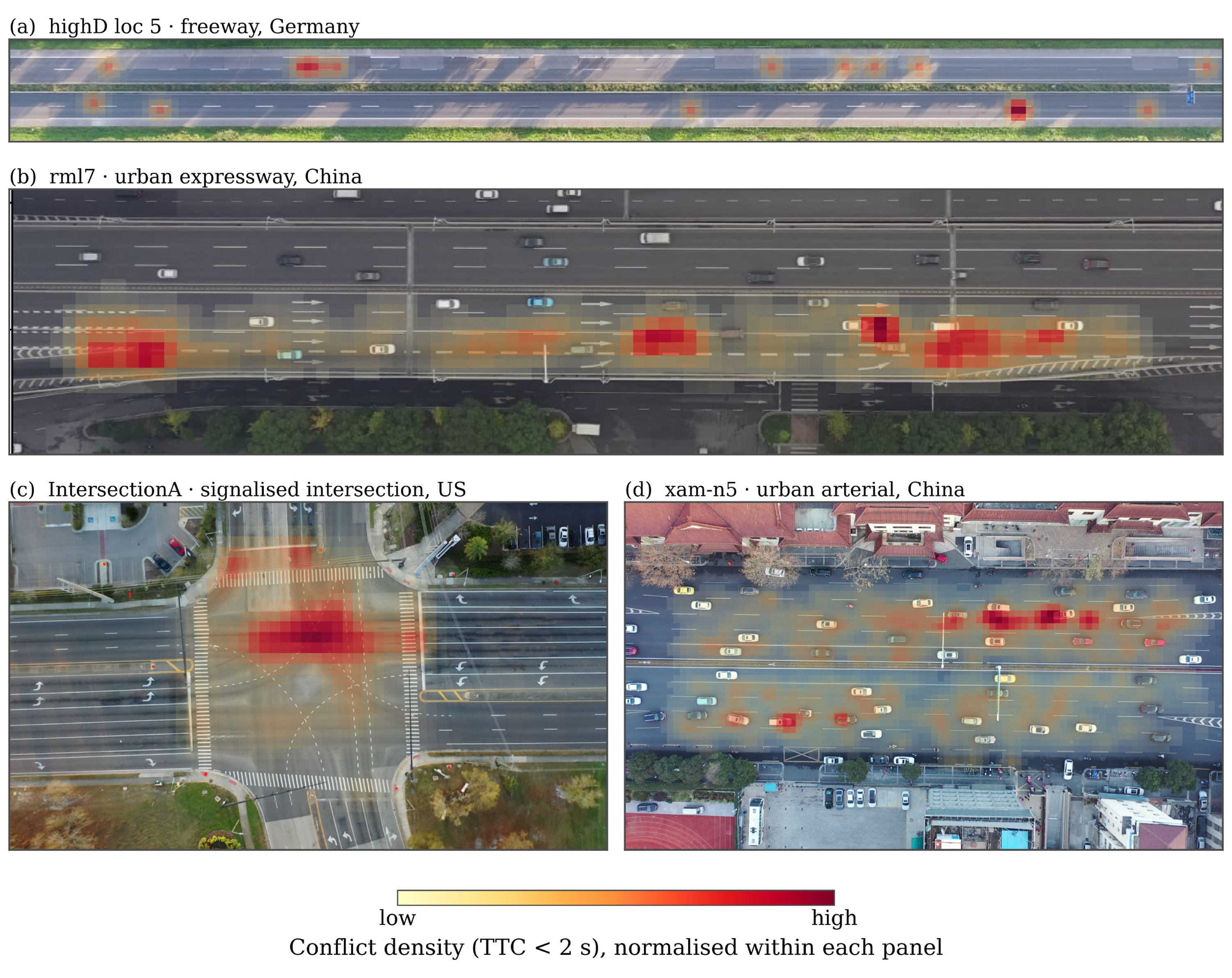}
  \caption{Density of identified conflict events at four Ozone sites.}
  \label{fig:conflict}
\end{figure}

Figure~\ref{fig:conflict} maps where the identified conflicts fall at four typical road sections. Figure~\ref{fig:conflict}(a) is a German freeway segment, Figure~\ref{fig:conflict}(b) a Chinese urban expressway, Figure~\ref{fig:conflict}(c) a signalized intersection in the United States, and Figure~\ref{fig:conflict}(d) a Chinese urban arterial. The shading gives the density of events with a time to collision below two seconds. On the freeway the events spread along both carriageways without a dominant location. On the urban expressway they concentrate in the lanes adjacent to the ramp connections, where vehicles entering and leaving the mainline force sharp adjustments. At the intersection they concentrate in the crossing and left-turn area. On the arterial they follow the through lanes in both directions. Color is normalized within each panel rather than across the figure, because the sites differ in observation duration and in bin size, so the shading is comparable within a panel but not between panels.

\subsection{Crash Data and Reconstruction}

The crash resources cover human-driven vehicles, automated driving systems (ADS), and advanced driver assistance systems (ADAS). Official human-driving records include sources from the United States, Great Britain, European Union countries, and Hong Kong \citep{nhtsa2025fars, dft2025stats19, ec2025care, hktd2025accidents}. Regulatory reports distinguish ADS from Level 2 ADAS \citep{nhtsa2021sgo}, and instrumented field data provide additional driver-assistance observations. Source reporting criteria differ. For example, FARS covers fatal crashes, while STATS19 and CARE concern reported injury crashes. These records support the study of different driving modes, but raw counts alone cannot establish their relative safety.

The crash schema records agent attributes such as driving mode, vehicle type, and interaction role, together with collision type, road conditions, traffic control, and contributing factors. Source reports and reconstructed motion remain distinct forms of evidence. The former records reported facts, while the latter supplies a plausible sequence for trajectory analysis and simulation.

A crash report is a narrative rather than a trajectory. A report lists the collision type, the road surface, and the contributing factors, but it does not provide the continuous motion of each vehicle before impact \citep{wang2019crash}. We therefore reconstruct that motion. The procedure reads the report fields, the road geometry from the matching digital-twin map, and any scene description, and it produces a physically plausible multi-agent trajectory over the pre-crash and impact phases \citep{kumar2018novel}. Map geometry restricts the region in which a vehicle could have traveled, and vehicle motion limits constrain changes in speed and direction. These constraints reduce the set of plausible sequences but do not uniquely recover unobserved motion. Reconstructed tracks should therefore be interpreted as report-consistent scenarios rather than measured ground truth.

\begin{figure}[tbp]
  \centering
  \includegraphics[width=\linewidth]{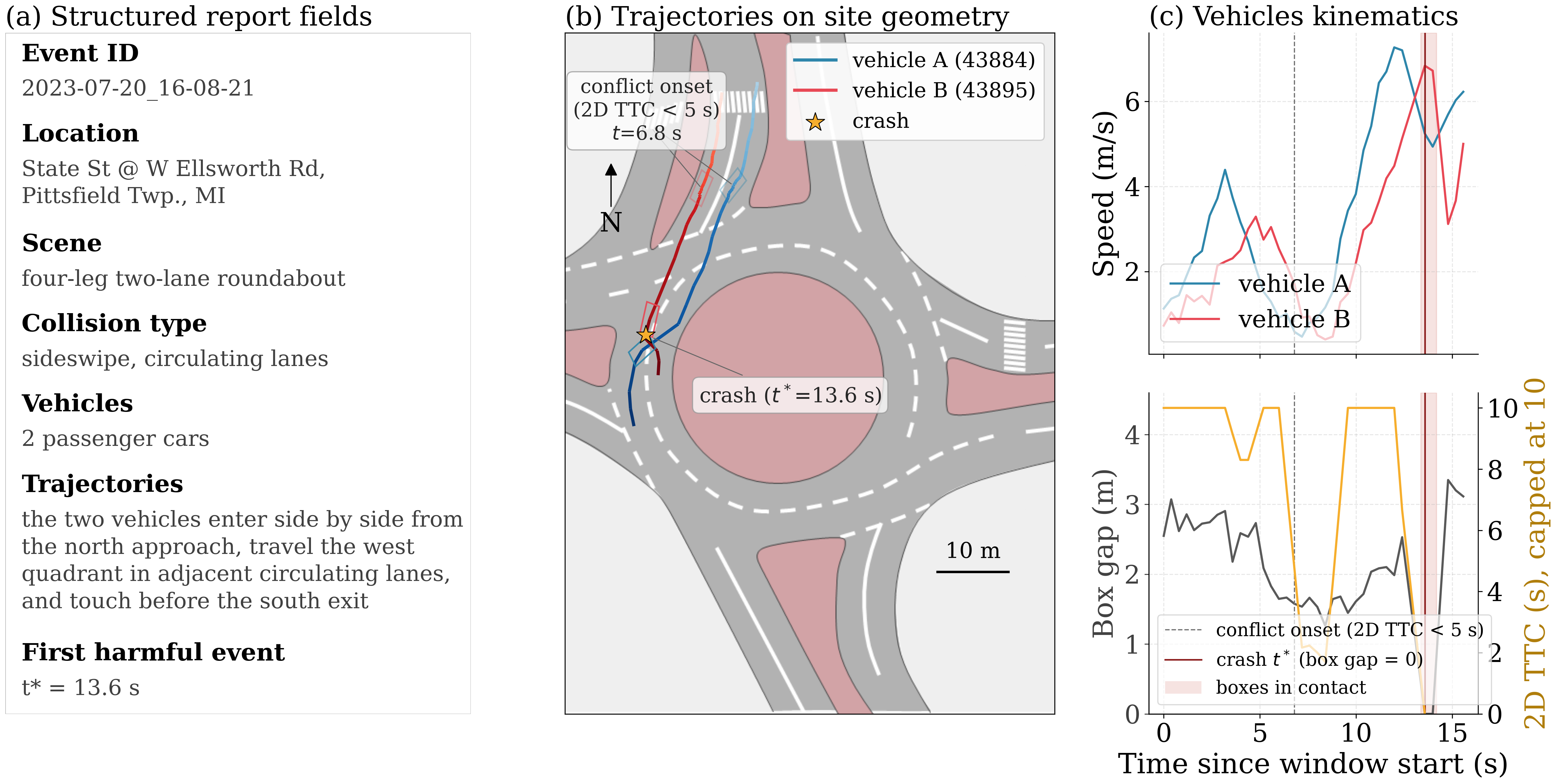}
  \caption{A crash event linked to vehicle-level motion under the Ozone schema.}
  \label{fig:crash}
\end{figure}

Figure~\ref{fig:crash} shows a crash record under the schema, using a sideswipe between two circulating vehicles at a roundabout site. Figure~\ref{fig:crash}(a) lists the structured fields of the event record, which give the event identifier, the location, the scene type, the collision type, the vehicles involved, a description of their paths, and the time of the first harmful event. Figure~\ref{fig:crash}(b) draws the trajectories of the two vehicles on the site geometry with a time gradient, together with their oriented bounding boxes at the conflict onset and at the crash moment. Figure~\ref{fig:crash}(c) gives the speed profiles, the bounding-box gap between the vehicles, and the box-based two-dimensional time to collision, which reaches zero when the two boxes touch. The two-dimensional time to collision falls below the five-second threshold 6.8 seconds into the window and contact occurs at 13.6 seconds, so the surrogate measure marks the interaction 6.8 seconds before impact. The event links its structured fields to the motion of the two vehicles involved, so the moments leading to the impact can be read in the same terms as any other trajectory in Ozone, including the surrogate safety measures evaluated over the event window.

\subsection{Digital-Twin Maps}

\begin{figure}[tbp]
	\centering
	\includegraphics[width=\linewidth]{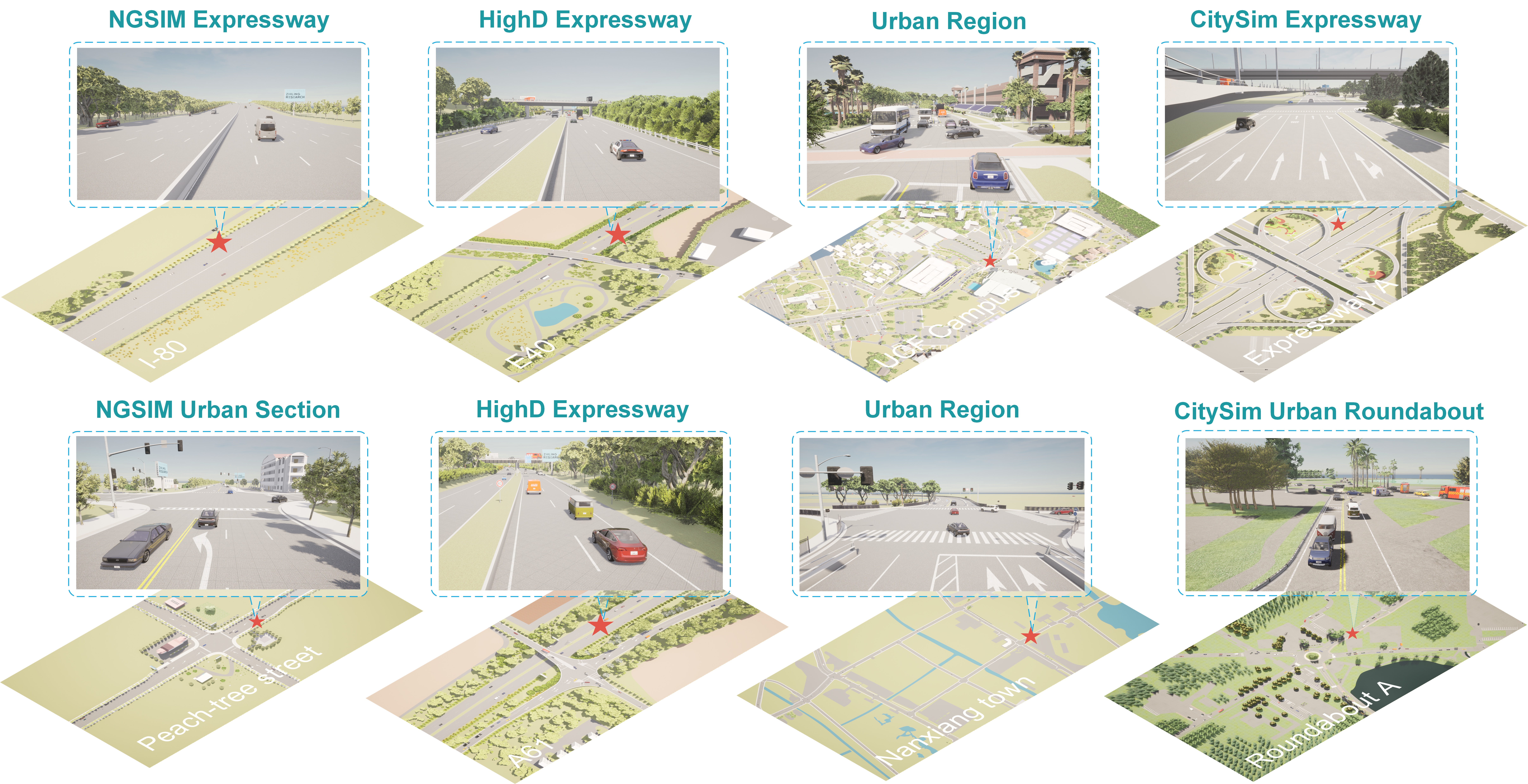}
	\caption{Digital-twin maps for eight Ozone sites, each pairing an isometric view of the reconstructed road with a rendered ground-level view.}
	\label{fig:map}
\end{figure}

The digital-twin component provides three-dimensional CARLA maps that reproduce road geometry, lane markings, traffic signals, and principal roadside features at the corresponding sites \citep{dosovitskiy2017carla}. The maps are built from aerial imagery, elevation data, and road vectors from OpenStreetMap, and they are refined with point-cloud surveys and street-level references so that the virtual road matches the real one \citep{zheng2023citysim}. Distributing the maps with the data removes the map-building step that co-simulation would otherwise require at every site. Figure~\ref{fig:map} shows eight of them. Each entry pairs an isometric view of the reconstructed site with a rendered ground-level view taken at the starred point, covering the NGSIM I-80 freeway and Peachtree Street corridors, the highD E40 and A61 autobahn segments, urban regions at the UCF campus and in Nanxiang, and the CitySim Expressway A and Roundabout A sites.

\begin{figure}[tbp]
	\centering
	\includegraphics[width=0.8\linewidth]{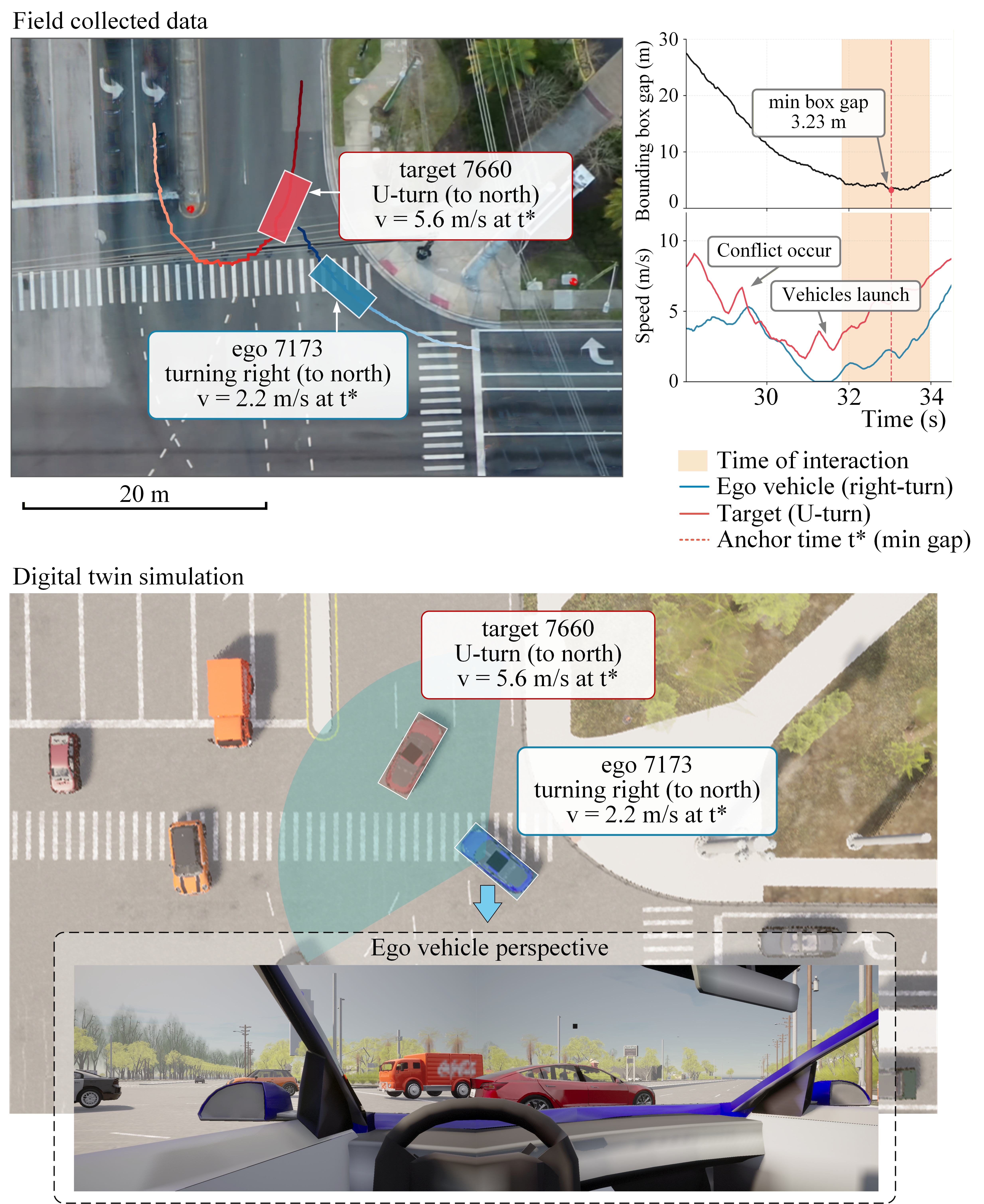}
	\caption{A digital-twin map creates additional viewpoints for field-collected bird's-eye data.}
	\label{fig:conflict_case}
\end{figure}

Digital-twin maps extend the viewpoints available for a recorded interaction. A standardized scene replayed on its matched map can be rendered from additional viewpoints and through the sensor models supported by the simulator. An overhead conflict can acquire a driver's view, and an ego-centered log can acquire a plan view. These outputs are simulated observations whose quality depends on map, rendering, and sensor-model fidelity. Figure~\ref{fig:conflict_case} shows a right-turning vehicle and a vehicle making a U-turn at a signalized intersection. The field-collected data in the upper part give the bird's-eye trajectories, the bounding-box gap, and the speed profiles, with a minimum gap of 3.23 meters at the anchor time. The digital-twin simulation in the lower part reproduces the same interaction on the matched map and adds the view from inside the ego vehicle, which the original bird's-eye recording cannot provide. Co-simulation drives calibrated traffic through the same geometry by coupling the microscopic logic of SUMO with the rendering of CARLA, which allows perception and control to be tested under realistic scenes \citep{lopez2018sumo}. The maps also constrain crash reconstruction, because the road geometry fixes the region in which each vehicle could have traveled.

\subsection{NBDT Scenario Resources}\label{sec:nbdt}

The Next-generation Benchmark by Digital Twin (NBDT) organizes geographically anchored scenario resources within Ozone. Its purpose is to connect real-world motion and safety events to a matched simulation environment. Each scenario combines a two-dimensional bird's-eye site reference, standardized tracks, event annotations when available, and a corresponding CARLA map. This organization allows a recorded interaction to be studied as a trajectory sequence, replayed in three dimensions, or used to initialize a closed-loop experiment.

NBDT brings together UAV trajectories, vehicle-centered observations, and risky-event segments. The UAV component records road-user motion over intersections and corridor segments. The vehicle-centered component uses adaptation scripts for sources such as Waymo to retain the ego vehicle, neighboring agents, and map context under the same schema \citep{sun2020scalability}. The risky-event component links extracted conflicts and reconstructed crashes to the relevant agents and temporal markers. These are complementary views of a scenario rather than independent trajectory standards. Model development and evaluation can therefore reuse the same scene representation across traffic-flow, prediction, planning, and safety tasks.

\subsection{Geographically Attributed Literature Data}

Beyond motion and safety records, Ozone includes a geographically attributed autonomous-driving literature dataset. It links bibliographic information to publication venues, research topics, author affiliations, and study locations where these can be identified. These attributes support analyses of the spatial distribution of research activity and differences in research topics across regions.

Affiliation locations identify where research is produced, while study locations identify the environments to which a paper relates. Keeping these meanings separate is necessary when interpreting geographic patterns. The literature resource complements the trajectory and scenario components by connecting research evidence to institutional and regional context. It does not form part of the trajectory-converter or conflict-extraction coverage reported above.


\section{Sensing and Data Processing}\label{sec:sensing}

\subsection{Converting Existing Datasets}\label{sec:conversion}

Most public datasets are distributed as trajectories rather than as raw video, and for them the task is not extraction but reconciliation. Conversion involves more than reformatting, because sources differ in the conventions they encode and because the fields the schema requires are often absent. Conversion therefore proceeds in four dependent steps, followed by manual calibration.

The first step aligns coordinates and road geometry. Public datasets express position in pixels, in feet, in a local metric frame, or in global coordinates. Every position is transformed into a common local metric frame, global coordinates are retained when the source provides them, and the road geometry of the site, including lane centerlines and edges, is registered against the corresponding digital-twin map. The second step unifies object classes, because each dataset labels road users with its own taxonomy and some use numeric codes. Every source label is mapped onto a single class set covering cars, trucks, buses, motorcycles, and other users. Class mapping is used when the source does not provide vehicle dimensions and a class-based estimate is required to complete the box.

The third step completes the oriented bounding box. Some datasets provide only a center point, and others provide an axis-aligned box that ignores orientation. When the vehicle dimensions are known, a box of the correct length and width is placed and oriented by the heading. When only the center point is available, the dimensions are estimated from the object class and the box is oriented along the direction of motion. The fourth step completes the heading and the remaining motion fields. When heading is absent it is estimated from the velocity vector of the smoothed trajectory, and the last reliable heading is held through low-speed and stationary segments where the direction of motion is undefined. Missing speed and two-dimensional acceleration fields are computed by finite differences on the smoothed positions and filtered to suppress numerical noise. Existing source fields are reconciled with the coordinate and unit conventions during conversion.

The automatic steps are followed by manual calibration and quality assurance. Where heading must be estimated from motion, an initial estimate is obtained before orienting the box, and the final motion fields are checked after smoothing. For each site we verify the pixel-to-meter scale and the alignment of the trajectories against the basemap, and we correct any systematic offset before the conversion is finalized. We inspect a sample of completed bounding boxes against the vehicle appearance in the source imagery and check the heading and class assignments on the same sample. Tracks with implausible motion, such as sudden jumps or impossible accelerations, are flagged for review and are then corrected or removed.

\subsection{Collecting and Processing New Data}

Ozone also includes newly collected observations. These follow a separate extraction path before entering the same trajectory schema. 

\subsubsection{Sensing Devices and Platforms}
\begin{figure}[tbp]
\centering
\includegraphics[width=\linewidth]{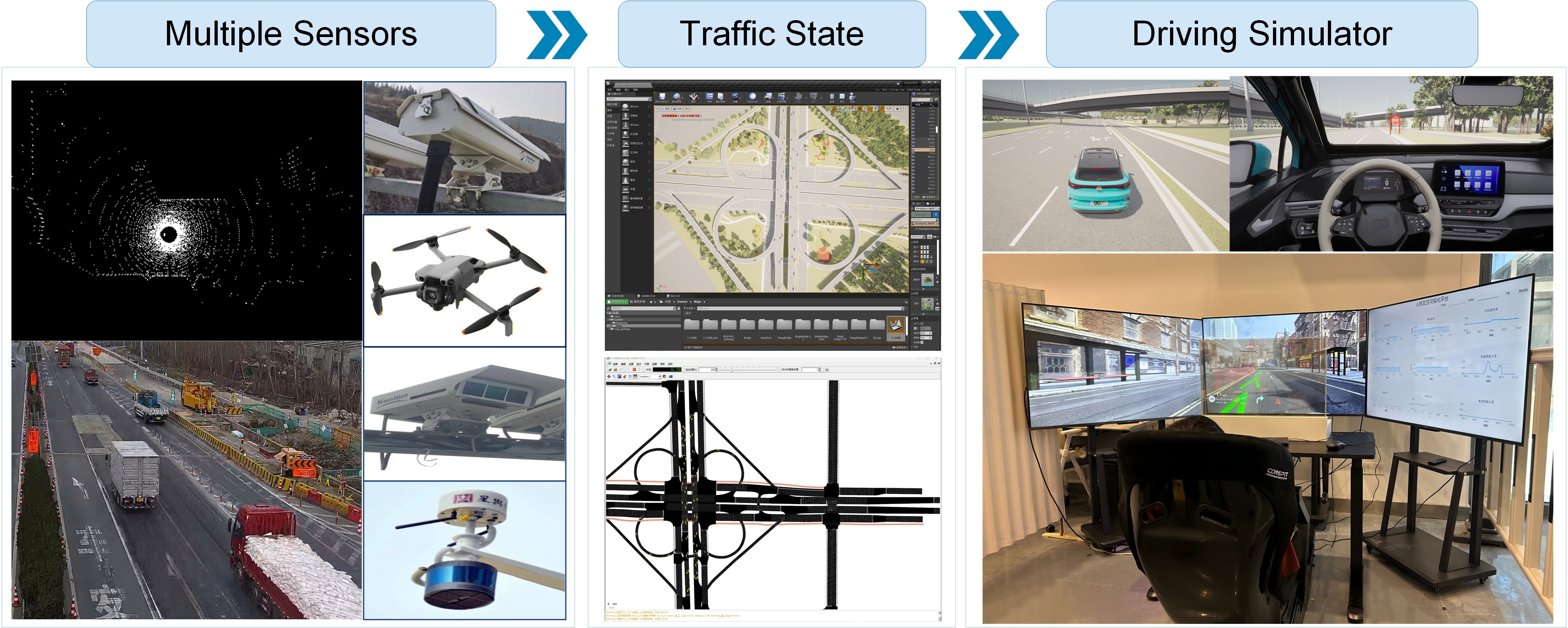}
\caption{Sensing and simulation setup behind the dataset.}
\label{fig:hardware}
\end{figure}

Ozone rests on the sensing devices and simulation platforms that Figure~\ref{fig:hardware} lays out from left to right. Multimodal sensing on the left supplies the road geometry and the traffic state, through roadside cameras, drones, roadside and vehicle LiDAR, and satellite positioning, alongside the point clouds and video frames those devices return. The digital-twin in the center reconstructs the site in three dimensions and carries the calibrated traffic state. Trajectories extracted from the sensors provide background traffic for scenario replay and simulator experiments. The driving simulator on the right renders a scene for human-computer interaction, which can be viewed from the driver's perspective or within the physical device.

Cameras remain the most widely deployed instrument, because they capture the state of people, vehicles, and the surrounding road with high spatial detail. Roadside cameras, usually mounted on signal poles, provide an elevated and continuous view of an intersection or an arterial, together with the ambient lighting and visibility conditions. Drone-borne cameras extend this with a hovering bird's-eye view over larger areas, which makes it practical to observe an entire merge or weaving segment at once without the occlusion that affects oblique roadside views \citep{zheng2023citysim, feng2025drone}. In-vehicle and dashboard cameras record the forward scene of the ego vehicle and the surrounding participants, and they also observe the driver, including gaze direction, posture, and attention level.

LiDAR and millimeter-wave radar complement camera sensing with direct depth measurements that are robust to lighting variation. Roadside LiDAR detects and tracks all road users within a volumetric field of view, and vehicle-mounted LiDAR supports on-board perception. Radar has lower spatial resolution but performs well in adverse weather and gives reliable range and velocity at longer distances. Satellite positioning contributes platoon tracks from controlled car-following experiments. Electroencephalography and eye tracking record the physiological signals that human-factor research requires, including cortical activity associated with workload and fatigue, and fixation and pupil dynamics associated with visual attention \citep{liu2025eeg, al-gburi2024driver}. When these streams are synchronized with the trajectory and the environment, the internal state of a driver can be related to external driving performance.

Each device produces data in its own coordinate frame, at its own resolution, and on its own temporal cadence. A roadside camera operates at 30 frames per second in pixel coordinates, whereas a LiDAR scanner outputs 10 Hz point clouds in a local Cartesian frame. Fusing these streams into one spatiotemporal representation requires calibration metadata that most public datasets do not record, including intrinsic and extrinsic parameters, synchronization timestamps, and georeferencing anchors. Ozone therefore stores the sensor type, the mounting geometry, the coordinate reference, and the sampling configuration alongside every record, so that the downstream pipeline can select the correct calibration and transformation automatically.

The hardware schema also covers simulation platforms that reproduce a site under controlled conditions. The simulator interface records scenario settings, vehicle states, control inputs, and synchronization with physiological streams. It supports both desktop simulators for participant studies and motion-platform simulators for immersive experiments. Driving simulators support human-in-the-loop experiments with configurable road geometry, traffic density, and weather. Co-simulation that couples the microscopic traffic logic of SUMO with the rendering engine of CARLA supports automated-vehicle testing with a realistic perception pipeline \citep{lopez2018sumo, dosovitskiy2017carla}. Because scenarios are generated from the digital-twin maps distributed with the data, a simulator experiment is grounded in a real road geometry and a calibrated traffic pattern rather than in a synthetic approximation of one.

\subsubsection{From Raw Video to Standardized Tracks}

For video observations, Ozone applies the four-stage pipeline shown in Figure~\ref{fig:pipeline}, comprising image stabilization, raw trajectory extraction, trajectory compensation, and trajectory standardization. Each stage operates on the output of the previous one, and together they turn raw video into analysis-ready tracks. Other sensing modalities require their own extraction procedures before standardization.

\begin{figure}[tbp]
\centering
\includegraphics[width=0.9\linewidth]{figures/data_process.jpg}
\caption{The Ozone processing pipeline.}
\label{fig:pipeline}
\end{figure}

\textbf{Image stabilization.} Wind and airflow can introduce unwanted camera motion into drone video. The stabilization stage detects feature points on fixed structures such as buildings and lane markings and matches them across frames with the Scale-Invariant Feature Transform \citep{opencv_sift}. Each frame is then warped to a reference frame by an affine transform, which cancels translational and rotational drift while preserving true motion. When feature matching fails, as it does when wind moves roadside vegetation, the stage falls back to a Channel and Spatial Reliability Tracker \citep{zheng2023citysim, feng2025drone}.

\textbf{Raw trajectory extraction.} Vehicles are detected with a Mask R-CNN model for instance segmentation, which returns an instance mask for each detected vehicle \citep{he2017mask}. A rotated bounding box is derived from the mask that follows the true shape and orientation of the vehicle, and it is therefore more suitable for safety analysis than an axis-aligned box, which overstates the occupied area of a turning or merging vehicle. The frame-level detections are then associated over time into raw multi-agent trajectories.

\textbf{Trajectory compensation.} The raw trajectories still carry error from viewpoint drift, imperfect stabilization, and occasional detector noise. The affine transforms estimated during stabilization are propagated to the detected positions, and the trajectories are smoothed to suppress jitter while preserving physically plausible motion. This stage matters most for video with mild but persistent camera motion, where residual drift would otherwise bias the position estimates.

\textbf{Trajectory standardization.} The final stage writes the trajectories into the schema. It assigns globally unique identifiers, computes heading and acceleration, transforms positions into a common local metric frame, and attaches the scene metadata and the map links.


\section{Model, Evaluation, and Prototype Interfaces}\label{sec:system}

The dataset in Section~\ref{sec:dataset} and the processing procedures in Section~\ref{sec:sensing} provide the inputs for the upper layers of Ozone. Model interfaces specify which trajectory, event, and scene fields a method consumes and which predictions or control signals it returns. Evaluation interfaces connect those outputs to metric suites and benchmark settings. Prototype interfaces carry the same data conventions into deployment-oriented applications.

The data schema is the common contract across these layers. A traffic-flow model reads standardized positions and motion fields, a safety model also reads interaction windows and surrogate measures, and a simulator uses the corresponding map and traffic-control context. Source-specific conversion remains within the data layer, so model and evaluation code can be reused across converted datasets. Derived fields still depend on the available source information, and each benchmark must select records with the fields it requires.

\subsection{Model Schema}\label{sec:models}

The model schema defines standardized interfaces for integrating computational models into the Ozone framework, supporting models ranging from classical traffic-flow formulations to state-of-the-art deep-learning architectures.

\subsubsection{Traffic Models}

The registry includes traffic-flow and safety models, alongside the AI models described below.

\emph{Traffic-flow models} include calibrated fundamental-diagram models (Greenshields, Greenberg, Underwood) and microsimulation models (IDM, MOBIL, Newell) that can be directly applied to any standardized trajectory dataset. 

\emph{Safety models} include crash-risk classifiers and conflict-detection modules that operate on the pre-computed surrogate safety measures.

\subsubsection{AI Models}

The AI interfaces support both traditional machine learning and deep learning.

\emph{Prediction models}. Trajectory-prediction baselines include recurrent neural networks, graph neural networks, and transformers. They accept standardized trajectory records and return predictions under the same input and output conventions. 

\subsubsection{World Foundation Models}
World models and foundation models connect tasks across the research layers of Ozone \citep{russell2025gaia, zheng2025world, hu2023gaia, wang2024drivedreamer}.
A \emph{world model} is a learned simulator of environment dynamics. Given a history of observations $o_{1:t}$ (e.g., camera frames, LiDAR sweeps, UAV bird's-eye views, or multi-agent trajectories) and a sequence of actions or scenario directives $a_{1:T}$, the model predicts plausible future observations $o_{t+1:T}$.
Following the formulation popularized by \citet{ha2018world} and the Dreamer series \citep{hafner2019dream, hafner2020mastering, hafner2023mastering}, a latent-state formulation uses a compact state $z_t$ to summarize the scene. A representative factorization is
\begin{align}
z_t  &\sim q_\phi(z_t \mid o_{\leq t},\, a_{<t})   && \text{(encoder)} \\
z_{\tau+1}   &\sim p_\theta(z_{\tau+1} \mid z_\tau,\, a_\tau,\, c)  && \text{(latent dynamics)} \\
\hat{o}_\tau &\sim p_\theta(o_\tau \mid z_\tau)                     && \text{(decoder)}
\end{align}
where $\tau \geq t$ indexes the rollout horizon and $c$ denotes auxiliary conditioning signals such as map geometry, weather, signal phases, or a high-level task specification.\footnote{We simplify the deterministic-stochastic split $(h_t, z_t)$ of the Recurrent State-Space Model \citep{hafner2023mastering} into a single latent $z_t$ for presentation, and extend the dynamics with an auxiliary conditioning $c$ to cover controllable world foundation models such as GAIA-2 and Cosmos.}
At training time, $z_t$ is sampled from the encoder so that the decoder reconstructs the observed frame $o_t$. At inference time, the latent dynamics roll forward autoregressively from $z_t$ to produce $z_{t+1}, z_{t+2}, \ldots$, and the decoder renders each imagined latent back into a predicted observation $\hat{o}_\tau$.
Training maximizes an evidence lower bound that balances observation reconstruction against predictive consistency in the latent space, or, in diffusion-based variants, denoises corrupted future observations conditioned on the past and on $c$ \citep{brooks2024video, bruce2024genie}.
Once trained, the model can be rolled forward over a selected horizon. An agent ``imagines'' futures inside the latent space, which supports planning, counterfactual reasoning, and policy learning without additional on-road data collection \citep{hafner2023mastering}.

A \emph{world foundation model} (WFM) extends this idea to the foundation-model regime.
Rather than training a task-specific simulator per dataset, a WFM pretrains a single high-capacity network on heterogeneous, large-scale mobility data, typically hundreds of hours of multi-view video paired with trajectories, map context, and traffic-control logs, so that the learned dynamics $p_\theta$ generalize across cities, weather regimes, and downstream tasks.
This paradigm is being pursued in both industry and open academic ecosystems.
General-purpose physical-AI platforms such as NVIDIA Cosmos\footnote{\url{https://www.nvidia.com/en-us/ai/cosmos/}} provide large pretrained backbones that can be post-trained for traffic-specific synthesis, as exemplified by Cosmos-Drive-Dreams\footnote{\url{https://research.nvidia.com/labs/toronto-ai/cosmos_drive_dreams/}}, which generates controllable, multi-view, spatiotemporally consistent driving videos from HD-map and text conditions.
Representative driving-domain systems include GAIA-1/GAIA-2\footnote{\url{https://wayve.ai/science/gaia-2/}} for controllable multi-view scene generation \citep{hu2023gaia, russell2025gaia}, Vista\footnote{\url{https://github.com/OpenDriveLab/Vista}} for high-fidelity scene-level video synthesis \citep{gao2024vista}, the DriveDreamer family\footnote{\url{https://github.com/BraveGroup/Drive-WM}} for action-conditioned video generation \citep{wang2024drivedreamer, zhao2024drivedreamer}, OccWorld\footnote{\url{https://github.com/wzzheng/OccWorld}} for 3D occupancy forecasting \citep{zheng2024occworld}, DriveWorld for 4D scene-understanding pre-training \citep{min2024driveworld}, UniSim for interactive real-world simulation \citep{yang2024learning}, and World4Drive, which couples a latent physical world model with an end-to-end planner \citep{zheng2025world}.
Community-maintained toolkits such as OpenDWM\footnote{\url{https://github.com/SenseTime-FVG/OpenDWM}} further package training, inference, and standardized evaluation of driving WFMs on public benchmarks.
These approaches aim to adapt a pretrained backbone to multiple transportation tasks through fine-tuning or task-specific conditioning.

WFMs are particularly valuable for transportation research because they cut across all four core layers of Ozone rather than serving a single application.
At the \emph{data layer}, WFMs can generate additional scenarios. Safety-critical events such as cut-ins, emergency braking, and pedestrian-vehicle conflicts are long-tail by nature, and learning a simulator that can \emph{generate} rare scenarios is more scalable than waiting to \emph{observe} them on the road \citep{feng2023dense, ruan2024traffic}.
Ozone's crash-reconstruction pipeline also infers unobserved motion from report attributes under map and kinematic constraints (Section~\ref{sec:dataset}). World models could extend this process by generating alternative trajectories consistent with the same conditions.
At the \emph{model layer}, WFMs serve as generalizable backbones alongside the Safety Foundation Model and the neuromorphic reaction-time model included in the model schema, providing shared representations that can be fine-tuned for trajectory prediction, behavior classification, conflict detection, and risk forecasting without retraining a bespoke architecture per task.
At the \emph{evaluation layer}, they enable \emph{counterfactual reasoning}, ``what if the lead vehicle had braked 0.5\,s earlier?'' or ``how would the conflict have unfolded in rain?'', which supports controlled safety analysis when such changes cannot be staged on the road. Such counterfactuals can be scored with Ozone's pre-computed SSMs to yield quantitative safety comparisons.
At the \emph{prototype layer}, WFMs narrow the sim-to-real gap by grounding synthetic futures in data statistics rather than in hand-tuned behavioral rules, supporting closed-loop validation for AV planners, roadside-unit safety monitors, and adaptive signal controllers (Section~\ref{sec:prototype}).
Human-factor research (Section~\ref{sec:human_factors}) can also use WFM-generated background traffic. This traffic can diversify driving-simulator scenarios beyond the fixed set replayed from CitySim, enabling larger-scale and more varied takeover or risk-perception experiments.

Prior efforts have partially addressed these needs.
ScenarioNet and its companion simulator MetaDrive\footnote{\url{https://metadriverse.github.io/scenarionet/}} \citep{li2023scenarionet} unify heterogeneous logs from Waymo, nuScenes, nuPlan, and Lyft into a common scenario description and support closed-loop replay, while CarDreamer\footnote{\url{https://github.com/ucd-dare/CarDreamer}} \citep{gao2024cardreamer} instantiates Dreamer-style world-model learning directly inside CARLA, demonstrating that latent-imagination policies can be trained end-to-end on urban driving tasks.
Within Ozone, RoadTailBench (Section~\ref{sec:roadtail}) already uses LLM-driven scenario compounding as an early form of generative scene synthesis, anticipating the tighter WFM integration supported by these interfaces.
Ozone supports WFM research through the following data and simulation resources.

\emph{Structured conditioning signals}. The trajectory schema specifies oriented bounding boxes, heading, speed, and two-dimensional acceleration aligned to a local metric frame, along with scene-level metadata such as weather, signal phases, and road type when available. These fields provide explicit conditioning signals for a WFM.

\emph{Matched rollout environments}. For a scene with a CARLA digital-twin map, lane geometry and traffic-control logic support replay and closed-loop execution of generated trajectories. The same scene can be used in software prototypes and human-in-the-loop simulator experiments.

\emph{Safety supervision}. Where the required fields are available, Ozone provides precomputed surrogate safety measures (TTC, 2D-TTC, PET, DRAC, CAI) at the frame level, which can serve both as auxiliary loss terms during WFM training and as structured evaluation signals at inference time.
These resources complement log-harmonization tools such as ScenarioNet and learning platforms such as CarDreamer. Ozone connects the shared data representation to safety measures, matched maps, and human-factor experiments within the wider research workflow.

Accordingly, Ozone's model schema accommodates world foundation models through two standardized interfaces that interoperate with every layer of the system.
The \emph{generation interface} consumes a context tuple $(o_{1:t}, \text{map}, \text{traffic-control}, c)$ drawn directly from the data schema and returns generated observations or trajectories, allowing autoregressive and diffusion-based models to share a common output contract. Autoregressive trajectory generation is illustrated by MotionLM \citep{seff2023motionlm}. These outputs can supply downstream training data, simulator scenarios, and prototype test cases.
The \emph{evaluation interface} groups metrics at three levels. Trajectory accuracy is assessed with minimum average displacement error (minADE) and minimum final displacement error (minFDE) \citep{shi2024mtr, seff2023motionlm}. Data-driven simulators such as Waymax provide a related setting for rollout evaluation\footnote{\url{https://github.com/waymo-research/waymax}} \citep{waymax}. Scene realism is assessed by comparing speed, headway, turning-radius, and fundamental-diagram distributions between generated and observed rollouts. Safety consistency is assessed through SSM distributions and collision rates \citep{ruan2024traffic, li2025aigc}.
Combined with closed-loop replay in the digital-twin maps, this unified evaluation lets WFMs trained on Ozone be compared not only on how realistic their scenes look, but on whether they preserve the statistical, behavioral, and safety-critical structure of real traffic.
A curated index of the broader open WFM landscape is maintained by the community.\footnote{See e.g.\ \url{https://github.com/LMD0311/Awesome-World-Model} and \url{https://github.com/HaoranZhuExplorer/World-Models-Autonomous-Driving-Survey}.}

\subsubsection{Neuromorphic Computing} This brain-inspired model is built around simulating the coordinated operation of human brain functional regions. It maps functions such as perception, cognition, memory, and motor control into computational units, and characterizes how multimodal stimuli, including visual, auditory, and tactile signals, are transmitted, integrated, and processed across different brain regions, thereby predicting human reaction time under varying task loads, environmental complexity, and information-presentation conditions. Its key advantage is that it can not only predict reaction time outcomes, but also explain why responses become faster or slower in terms of the specific functional regions involved, the processing pathways selected, and the mechanisms of resource competition. For industry, this type of model has broad value in the design and optimization of intelligent vehicles, aviation warning systems, industrial control, wearable devices, and other human-machine interaction systems. It can help companies evaluate design alternatives through simulation at an early stage of product development, reduce reliance on large-scale human-subject experiments, lower R\&D costs, and improve system safety, interpretability, and cross-scenario applicability.

\subsubsection{Safety Foundation Models}\label{sec:safety_model} Based on Ozone's data, we trained a multimodal Safety Foundation Model to perceive and evaluate risks in complex traffic environments. The model integrates heterogeneous inputs, including camera imagery, trajectory data, and traffic signal information, to construct a unified spatiotemporal representation of dynamic road scenes. Through large-scale pretraining and cross-modal alignment, it captures interaction patterns among traffic participants and produces structured outputs describing risk understanding and conflict severity, enabling fine-grained analysis of vehicle-vehicle and vehicle-pedestrian conflicts.

Each model in the registry conforms to a unified interface specifying input schema (which fields from the trajectory/event data it requires), output schema (predictions, classifications, or control signals it produces), and evaluation hooks (which metrics to compute automatically upon inference). This design enables researchers to swap models within an experimental pipeline without modifying the surrounding data-loading or evaluation code.

\subsection{Evaluation Schema}\label{sec:evaluation}

The evaluation schema provides standardized protocols for measuring and comparing research outcomes across all layers of the system. It defines metric suites, benchmark configurations, and result repositories.

\emph{Metric suites} group related evaluation metrics by research domain. The safety-analysis suite includes all SSMs (TTC, PET, DRAC, etc.) with standardized computation procedures and threshold definitions. The trajectory-prediction suite includes ADE, FDE, and collision-rate metrics with specified prediction horizons. The traffic-flow suite includes RMSE and correlation metrics for fundamental-diagram calibration.

\emph{Benchmark configurations} specify fixed experimental setups for fair comparison. These include training, validation, and test splits, preprocessing procedures, hyperparameter search protocols, and computational budgets. Each benchmark is versioned, so that results remain comparable as the system evolves.

\emph{Result repositories} store published baseline results for each benchmark, enabling researchers to immediately compare their methods against established performance levels without re-running baselines.

\subsection{Prototype Schema}\label{sec:prototype}

The prototype schema extends the system toward deployment-oriented applications, bridging the gap between research outputs and real-world systems.

\subsubsection{Hardware Prototypes}
Hardware prototypes include autonomous vehicle platforms, connected-vehicle roadside units (RSUs), and adaptive signal controllers. Ozone defines standard data interfaces for each hardware type, enabling researchers to deploy algorithms developed within the system directly onto prototype hardware. For example, a safety-monitoring algorithm developed and validated using Ozone's data and evaluation schemas can be packaged for deployment on an RSU with minimal interface adaptation.

\subsubsection{Software Prototypes}
Software prototypes include city-scale traffic management dashboards, real-time safety monitoring systems, and autonomous-driving decision modules. The prototype schema specifies deployment interfaces, data-stream formats, and performance-monitoring hooks that maintain traceability from research-stage evaluation to deployment-stage operation.

\subsubsection{Deployment Criteria}
Each prototype category is associated with deployment criteria that define minimum performance thresholds, safety requirements, and validation procedures that must be satisfied before field deployment. These criteria draw directly from the evaluation schema, ensuring that the same metrics used to assess research-stage performance also govern deployment readiness.


\section{Case Studies}\label{sec:cases}

The platform provides a plug-and-play foundation for a broad range of transportation studies and substantially reduces the engineering overhead of building an end-to-end experimental pipeline. It supports the full workflow, from multimodal data organization and standardized benchmarking to model development and deployment-oriented validation. It can serve human-factor experiments through configurable driving-simulation scenarios, provide reproducible validation environments for assisted and automated driving algorithms, and support transferable evaluation across heterogeneous datasets and cities. The following case studies illustrate how the framework accelerates development and improves reproducibility.

\subsection{Human-Factor Research}\label{sec:human_factors}

Driving-behavior research requires observing perception, decision-making, and control in an environment that is both controllable and realistic. Real-road experiments are expensive and risky, especially in critical situations, so driving simulation has become a core platform. In such studies, datasets are essential for scenario construction, particularly for generating background traffic. Ozone provides simulation maps, reconstructed trajectories, and calibrated traffic annotations, which allows a simulator to reproduce traffic flows close to real operations and improves the realism, repeatability, and comparability of human-factor experiments.

\subsubsection{Problem Overview}

This case study addresses the risk-perception bias of drivers during emergency takeovers in automated driving. Prior work shows that when a Level 3 system issues a takeover request at a critical moment, drivers may miss hazards, mislocalize risk, or react too aggressively, and the takeover then fails. To mitigate this, we propose a risk-prompting scheme delivered through an augmented-reality head-up display (AR-HUD). A graph attention network with long short-term memory (GAT-LSTM) predicts the subjective attention of the driver and compares it with the objective risk distribution to identify overlooked hazards, which are then highlighted in the field of view with red three-dimensional wireframe cues. The method is evaluated in a simulated emergency-takeover setting.

\subsubsection{Experiment Benchmark}

We conducted an emergency-takeover experiment in a driving simulator. Scenarios and background traffic were generated from CitySim data, covering CitySim Intersections A, B, D, and E and Freeway D, and including common high-risk events such as sudden braking by the lead vehicle, cut-ins, crossing-traffic interference, and disturbance from complex flow. All scenarios used consistent daytime weather to avoid a weather confound. Ten representative takeover scenarios were designed, and ten drivers participated in a comparison with and without the display. Eye-tracking data were collected during each trial. The model estimated subjective attention, which was compared with objective risk to detect missed hazards, and these hazards were then presented through the display. Background traffic remained identical across conditions. The metrics were lateral stability, measured by the standard deviation of the heading angle, longitudinal stability, measured by the standard deviation of speed, and takeover success rate, measured as successful takeovers divided by all attempts. Results were averaged across the ten scenarios and participants.

\subsubsection{Results}

The scheme improves emergency-takeover performance. Relative to the condition without the display, the standard deviation of the heading angle decreased by 22.02\% ($p = 0.014$), which indicates better lateral stability. The standard deviation of speed decreased by 6.47\% ($p = 0.048$), which indicates smoother longitudinal control. The takeover success rate increased by approximately 8\% ($p = 0.023$). These findings provide a baseline for the design of takeover-assistance interfaces, for risk-perception correction, and for comparative driver-behavior studies.

\subsection{Automated-Driving Testing}\label{sec:roadtail}

Automated driving systems remain vulnerable to long-tail safety scenarios. Existing studies emphasize random traffic variation or extreme weather, and they under-address the structural long-tail risk caused by road-engineering hazards. Real-road testing of such hazards is dangerous and costly, which makes high-fidelity simulation essential. RoadTailBench addresses this need by combining co-simulation maps with realistic traffic flows derived from large-scale inspection records, and by covering compound engineering-defect scenarios with dynamic participants and adverse weather.

\subsubsection{Problem Overview}

A key difficulty in automated-driving safety is the robust semantic understanding of road infrastructure and safe decision-making under infrastructural ambiguity. This case study focuses on the decision failures caused by compound scenarios that combine road-engineering hazards, extreme weather, and dynamic participants. Typical examples include a speed-limit sign masked by sun glare, a low-visibility rainy curve, and misleading markings, each of which can induce a decision that appears reasonable but is dangerous. We convert real-world hazard scenes into high-fidelity parameterized simulation cases and quantify the perception blind spots and decision boundaries of mainstream benchmark algorithms.

\subsubsection{Experiment Benchmark}

RoadTailBench provides a standardized benchmark suite derived from real road hazards. It includes high-fidelity static road topologies in the ASAM OpenDRIVE format and dynamic scene elements in the ASAM OpenSCENARIO format, which together form a dedicated environment for structural long-tail risk testing. The scenarios replicate the physical hazard and add adverse lighting, weather, and occlusion, which improves edge-case coverage. Representative compound-defect scenarios are imported into CARLA, several benchmark algorithms with different control strategies are evaluated, and the ego vehicle is required to traverse the long-tail section without prior warning while the test system records interaction states and computes the predefined metrics.

\subsubsection{Results}

We developed a long-tail scenario library with 125 fully parameterized test cases, including targeted ablation cases for individual hazard elements. The library spans four operational design domains, namely urban, mountainous, suburban, and highway, and is grounded in 14 core categories of road-engineering hazard, such as misleading signs, severely worn lane markings, missing guardrails, limited sight distance on curves, and clearance intrusions. Language-model agents were used for cross-derivation and compound-risk superposition, which increases the scarcity and the testing value of the scenarios. Relative to a single static hazard, the generated variants introduce multi-factor interference, including twilight glare, dense fog, localized heavy rain, occlusion by heavy vehicles, aggressive merging, and pedestrians emerging from blind spots. This risk matrix increases the difficulty of decision-making and collision avoidance while remaining compatible with existing simulators, which enables direct robustness evaluation of benchmark algorithms.


\section{Results and Discussion}\label{sec:results}

This section synthesizes the outcomes of the platform against its design objectives and discusses the implications for transportation research.

\subsection{System Design Evaluation}

We evaluated Ozone through a structured assessment of conversion accuracy, schema completeness, and workflow efficiency. For conversion, we validated the automated pipelines by comparing the converted trajectory fields, namely position, heading, speed, acceleration, and the oriented bounding-box corners, against available source annotations for a random sample of 500 vehicles across NGSIM, highD, CitySim, and UTE. This validation sample covers four sources rather than all twelve completed converters. The mean positional error after conversion was below 0.15 m, the mean heading error below 2.5 degrees, and the mean speed error below 0.3 m/s, which indicates that conversion preserves the corresponding source fields within these errors. Agreement with source annotations measures conversion fidelity and does not establish the absolute accuracy of the original observations.

Schema completeness was assessed by cataloguing the fields required for downstream tasks, including trajectory modeling, safety analysis, and digital-twin construction, and by measuring the coverage ratio for each converted dataset. After conversion the four evaluated datasets reached complete coverage of the mandatory fields used in this assessment, whereas before conversion the coverage ranged from 45 percent for NGSIM, which lacks the oriented bounding box and heading, to 82 percent for CitySim, which provides most fields natively. Field completion includes derived quantities and does not mean that every field was directly observed. Engineering effort was assessed separately. In a study with five graduate researchers, the time required to set up a cross-dataset experiment fell from an average of 12.4 days with manual preprocessing to 1.8 days with the Ozone pipelines, a reduction of 85 percent.

\subsection{Unified Benchmark and Metrics}

The evaluation schema provides benchmark configurations that support fair comparison and direct reuse across studies, and we summarize the outcomes in three respects.

For safety analysis, Ozone provides a standardized pipeline that computes the surrogate safety measures in one pass. This removes the repeated re-implementation of metric scripts and makes safety results comparable across datasets and tasks, because the same event definitions, threshold settings, and observation windows can be reused for event mining and downstream model evaluation. Cross-site counts still require attention to recording duration, traffic exposure, and source coverage.

For behavior research, Ozone provides reusable digital-twin maps, standardized trajectories, and synchronized multimodal interfaces, which together build a controlled yet realistic experimental environment. Scenario families, participant protocols, and intervention settings are configured under one schema and evaluated with consistent metrics, which supports reproducible comparison across tasks such as risk perception, attention allocation, decision-making, and control stability.

For automated-driving safety testing, Ozone provides a path from the definition of a risk scenario to a simulator-executable benchmark suite. Structural road risks, environmental disturbances, and interactive traffic conditions are combined through standardized scenario specifications and evaluated under common logging and metric protocols, which enables a fair robustness assessment of different perception, planning, and control stacks.

\subsection{Reproducibility}

Reproducibility is a persistent difficulty in transportation research, where results often depend on undocumented preprocessing decisions, dataset-specific code, and evaluation settings. Ozone addresses this at three levels. Data reproducibility follows from the standardized schema and the automated converters, which ensure that any researcher starting from the same raw dataset obtains identical processed data. Methodological reproducibility follows from the standardized model interfaces and pre-implemented baselines, which allow published results to be reproduced by running the same model code on the same benchmark configuration. Evaluative reproducibility follows from fixed metric definitions, dataset splits, and evaluation scripts, which make reported numbers directly comparable across papers.

To quantify the improvement, we attempted to reproduce the results of five published trajectory-prediction studies, first with the original code and preprocessing of the authors, and then with the same model architectures retrained on the Ozone pipeline. In the first case only two of the five studies yielded results within 5 percent of the published numbers, and the remaining three showed discrepancies of 8 to 22 percent that arose from undocumented preprocessing variation. In the second case the standardized-pipeline experiments reported agreement within 3 percent across the five reproduced studies.

\subsection{Cross-City Transferability}

A central design goal is to allow findings to transfer across regions. The standardized representation normalizes away city-specific conventions, including coordinate systems, data formats, and naming conventions, while preserving the semantic content, including vehicle dynamics, traffic patterns, and safety events.

We evaluated cross-city transferability with the safety foundation model described in Section~\ref{sec:models}. The model was trained on data from two cities and evaluated on held-out data from a third, cycling through all three-city combinations. The average cross-city F1 score for conflict-type classification was 0.83, against 0.91 within city, a transfer efficiency of 91 percent. For crash-risk prediction the cross-city area under the curve was 0.79, against 0.86 within city, a transfer efficiency of 92 percent. The ratios express cross-city performance relative to within-city performance. These results indicate useful transfer in the evaluated settings, while standardizing the representation does not remove differences in traffic behavior, road geometry, or sensing quality between cities.

\subsection{Scalability}

The system is designed to accommodate growing data volumes and new sources. The resource collection totals approximately 2.3\,TB. The inventory contains 19 trajectory sources, of which twelve have completed converters. Conflict subsets are derived only where the required fields are available, and the digital-twin library provides the maps illustrated in Section~\ref{sec:dataset}. The source, converter, conflict, and map counts therefore describe different aspects of coverage. For a source with an implemented converter, processing requires approximately four to eight hours of computation, depending on source size and field completeness. This estimate excludes converter development.

The schema is extensible. A new sensor type is added to the hardware schema by specifying its output format and calibration parameters. A new data type, such as a vehicle-to-everything communication log or a weather-sensor stream, is integrated by defining its fields and its alignment rules. A new model is registered by implementing the standardized input and output interface. This modularity allows the system to grow without structural redesign.

\subsection{Scope and Limitations}

The common schema reduces differences in representation, but it does not make all sources equally informative. Class-based vehicle dimensions and motion-derived heading introduce uncertainty, especially during low-speed maneuvers. Safety measures remain sensitive to these fields and to the assumed motion model. A dataset with a completed converter may therefore remain unsuitable for a particular conflict measure. Similarly, a constrained crash reconstruction supplies a plausible scenario rather than a unique account of unobserved pre-crash motion.

Geographic coverage also differs across resource types. The world map includes source inventories and crash-report jurisdictions, while the ready-to-use trajectory subset and illustrated digital-twin library have narrower coverage. Crash reporting criteria differ between jurisdictions and driving modes, which prevents raw counts from serving as direct risk comparisons. The conversion assessment and workflow study cover four datasets and five researchers, respectively. The reported reproducibility and transfer results describe their evaluated settings and require broader validation before they can be generalized to every source and task.


\section{Conclusion and Future Work}

Ozone connects standardized transportation data to model development, evaluation, and deployment-oriented applications. Four core research layers cover Hardware, Data, Model, and Evaluation, while a Prototype layer extends the workflow toward use in hardware and software systems. The data layer links trajectories, traffic conflicts, reconstructed crashes, and digital-twin maps under one schema. The source inventory spans 19 trajectory datasets at 59 sites in eight countries, with completed converters for twelve sources and seven still being adapted.

The shared representation supports cross-dataset traffic modeling, safety analysis, trajectory prediction, and scene generation. Matched maps and scenario resources connect those records to human-factor experiments and automated-driving tests. Validation on four datasets reports mean conversion errors below 0.15\,m in position, 2.5$^\circ$ in heading, and 0.3\,m/s in speed. The workflow assessment reports an 85\% reduction in experiment setup time. These results support the value of shared data and interfaces within the evaluated settings, while source quality, inferred fields, and regional differences remain relevant to downstream conclusions.

Future work will complete the remaining converters and expand the digital-twin library to additional cities and road types, including rural roads and construction zones. Integrating real-time vehicle-to-everything and roadside-sensor streams will support cooperative driving and traffic management. Further work on pretrained foundation models will examine adaptation across cities and traffic conditions. Formal verification within the evaluation layer and community-maintained versioning and contribution procedures will support deployment-oriented validation and continued development of the platform.

\section{Data and Code Availability}

The Ozone platform, including standardized datasets, conversion pipelines, digital-twin maps, model baselines, and evaluation scripts, is publicly available at \url{https://ozone.zhilingtech.com/}. Source datasets remain subject to their respective access and reuse conditions.


\bibliographystyle{abbrvnat}
\bibliography{references}


\clearpage
\beginsupplement


\end{document}